**Enhancing Hot Carrier Collection for Solar Water Splitting with Plasmonic Titanium Nitride**

Alberto Naldoni[1,2], Urcan Guler[1], Zhuoxian Wang[1], Marcello Marelli[2], Francesco Malara[2], Xiangeng Meng[1], Alexander V. Kildishev[1], Alexandra Boltasseva[1], Vladimir M. Shalaev[1,*]

[1] School of Electrical & Computer Engineering and Birck Nanotechnology Center, Purdue University, West Lafayette, IN 47907, USA

[2] CNR-Istituto di Scienze e Tecnologie Molecolari, Via Golgi 19, 20133 Milan, Italy

The use of hot electrons generated from the decay of surface plasmons is a new paradigm to increase the conversion yield in solar energy technologies. Titanium nitride (TiN) is an emerging plasmonic ceramic that offers compatibility with CMOS technology, corrosion resistance, as well as mechanical strength and durability thus outperforming noble metals (i.e., Au, Ag) in terms of cost, mechanical, chemical and thermal stability. Here, we show that plasmonic TiN nanoparticles produce 25% higher photocurrent enhancement than Au nanoparticles decorated on $TiO_2$ nanowires for photoelectrochemical water splitting. Our results highlight that TiN offers superior performance in hot carrier generation due to enhanced absorption efficiency, increased electron mean free path, and the ability to form an Ohmic junction with $TiO_2$, thus enabling an extremely efficient electron collection not achievable with plasmonic Au nanoparticles. Our findings show that transition metal nitrides enable practical plasmonic devices with enhanced performance for solar energy conversion.

Photocatalysis is central to the development of a new clean energy economy. Photochemical reactions with semiconductors are driven by electrons and holes generated by sunlight absorption. Although photocatalysis is already used in small-scale abatement of both indoor and outdoor pollution, the low efficiency of solar-to-chemical energy conversion limits further breakthroughs in strategic applications such as hydrogen production and photovoltaics (PV). The engineering of composite interfaces for optimized operation is a key step to increase the solar absorption, electron-hole separation and collection, and finally to provide high reaction yield.

The use of plasmonic nanostructures in energy conversion applications has been demonstrated with the enhanced performance of solar-to-fuels and solar-to-electricity devices.[1–3] Surface plasmons concentrate light into nanoscale volumes at the interface with a semiconductor providing intense electromagnetic field localization and improved scattering. The possibility to harness hot electrons generated from the surface plasmon decay[4,5] has shown to be a promising route towards selective nanocatalysis,[6–9] full-spectrum solar water splitting,[10–12] and ultrafast photodetection.[13–16] The nonradiative decay of surface plasmons, occurring in the 40-150 fs timescale,[4] produces a population of highly energetic (*hot*) electrons that are stabilized through injection into the semiconductor conduction band across a Schottky barrier.[17] Plasmonic nanoparticles (NPs) enable the efficient conversion of solar light into electrons with an energy well below the band gap of the semiconductor but limited to energies higher than the potential barrier ($\phi_B$). As of yet, these demonstrations have been limited to NPs made of plasmonic noble metals such as Ag and Au. Despite their good optical performance, several challenges remain for noble metal NPs including high cost, poor chemical (Ag) and thermal stability (Au, Ag), diffusion into surrounding structures, and CMOS incompatibility that hinder the practical implementation of conventional plasmonic structures.

Aluminum nanocrystals are alternative plasmonic photocatalysts[9] that provide efficient hot electron generation at low cost (i.e., high abundance of Al on the earth crust). Nevertheless, major issues for large-scale use are related on Al nanoparticles preparation, due to explosive reactivity of Al molecular precursor with air and water, and poor chemical and thermal stability of final metallic Al nanostructures. On the other hand, transition metal nitrides such as titanium nitride (TiN) and zirconium nitride (ZrN) have recently been proposed as plasmonic materials that exhibit gold-competitive optical properties while also offering compatibility with CMOS technology, thermal and chemical stability, corrosion resistance, as well as improved mechanical strength and durability in comparison to noble metals[18,19]. For instance, TiN has been shown to exhibit superior performance in local heating, and in extremely high temperature applications such as heat-assisted magnetic recording and solar/thermophotovoltaics[20,21]. In addition, TiN has a work function of $\phi_M$~4 eV against the vacuum[22], which is much lower than the characteristic value reported for Au ($\phi_M$~5.2 eV)[23]. Since the work function of TiN is larger or equal to the electron affinity of most semiconductor metal oxides widely used in photocatalysis such as $TiO_2$, $\alpha$-$Fe_2O_3$, and also Si, TiN is expected to form favorable energetic alignment for hot carrier-enhanced solar energy conversion.

To demonstrate the distinct advantage of using TiN for the photoelectrochemical water splitting, we integrate plasmonic TiN and Au NPs in two separate $TiO_2$ nanowires (NWs) scaffolds. We show that the TiN/$TiO_2$ hybrid structure produces a 25% increased photocurrent in comparison to that of the Au/$TiO_2$ because TiN provides larger absorption efficiency and thus enhanced hot electron generation. We further demonstrate that TiN forms an Ohmic junction at the $TiO_2$ interface in contrast to the classical Schottky barrier formed in the case of Au/$TiO_2$, thereby allowing "downhill" hot electron collection into the semiconductor and enabling higher conversion efficiencies.

**Nano-building blocks for plasmonic photoelectrodes**

We developed a photoanode platform for the photoelectrochemical water splitting based on $TiO_2$ NWs and plasmonic NPs made of Au or TiN. Their surface plasmon resonance extend solar energy harvesting from 380-400 nm ($TiO_2$ band gap is 3.2 eV) to 600 nm and 800 nm and further in the near infrared (NIR) wavelength range, for Au and TiN NPs, respectively (Supplementary Figure 1).

The glancing angle deposition (GLAD) technique[24,25] was used to grow $TiO_2$ NW arrays on fluorine-doped tin oxide (FTO) substrates. Figure 1a shows that the fabricated $TiO_2$ NWs are uniform on a square centimeters scale (Supplementary Figure 2), which is required for highly efficient scalable devices. The NWs grow up from a narrower point (near the FTO substrate) and almost triple the width up to ~230 nm, while having an average length of 830 nm (Figure 1b). After annealing at 500 °C in an ambient atmosphere for 6 h, $TiO_2$ NWs crystallize in the pure anatase phase with near single crystalline quality (Figure 1c and Supplementary Figures 3-4) as shown by the FFT analysis and selected area electron diffraction (SAED). The crystalline indexing of FFT image highlights a general orientation along the [0,2,-1] zone axis next to several crystalline grains with [-1,1,1] orientation.

Next, separate $TiO_2$ NW arrays are functionalized with either Au or TiN NPs (with an average diameter of 20 or 50 nm, see Supplementary Figures 5-6). Nanoparticles deposited onto anatase NWs are well dispersed and attached to the NWs at different spots such as on the side, on the peak edge, or near the bottom of the NW (Figures 1 d-g). Figures 1f-g show a single TiN nanocube on the $TiO_2$ surface highlighted through the elemental maps obtained via filtered images at the characteristic electron energy loss (EEL) edges of Ti ($L_{2,3}$), O (K) and N (K). As discussed previously in more detail[26], TiN nanocubes show a 1-2 nm surface oxide layer (see also Supplementary Figures 7) containing $TiON$-$TiO_2$ species that enable a stronger adhesion than Au, and improved electronic interface coupling, on $TiO_2$ NWs. The native oxide at the surface of the plasmonic TiN nanoparticle is a unique and crucial characteristic for efficient hot electron injection in plasmonic material-semiconductor composites[27,28].

To compare the surface coverage of plasmonic NPs on $TiO_2$ NWs, we measure the photoanodes open circuit voltage ($V_{OC}$) decay. This method probes the quasi-Fermi level of the semiconductor in light and dark conditions. During decay transition processes, we obtain kinetic information on charge exchange between the electrode and the electrolyte (a detailed discussion can be found in the Supplementary Information). The charge recombination is mediated by surface trap-states and NPs deposition reduces their number on the $TiO_2$ surface, producing a decrease in the characteristic $V_{OC}$ decay time (half life time, $\tau$)[10]. From the fitting of the $V_{OC}$ curves (Supplementary Figure 8) we observe a similar decrease of $\tau$ from 0.54 s for $TiO_2$ to 0.15-0.20 s for $TiO_2$ –Au/TiN plasmonic electrodes that verifies a similar loading of Au and TiN NPs[10].

**Photoelectrochemical water splitting and optical properties**

We test the $TiO_2$ NW-based films in the photoelectrochemical water oxidation in a typical three-electrode cell. Figures 2a-b show the photocurrent density (J) as an anodic bias is applied to bare and NP-loaded $TiO_2$ NWs. Bare $TiO_2$ NWs produce 0.55 mA/cm$^2$ at 1.23 V vs. the reversible hydrogen electrode (RHE), while substantial increase of photocurrent is observed when plasmonic NPs are attached to the TiO$_2$ NWs. The electrode containing 50-nm plasmonic TiN NPs shows a photocurrent of 0.91 mA/cm$^2$, while the equivalent one containing 50-nm Au NPs exhibits 0.77 mA/cm$^2$. Consequently, the TiN-modified TiO$_2$ outperforms the Au-containing electrode, achieving a 25% photocurrent enhancement. 50-nm Au and TiN NPs produce higher enhancement (+40% and +65%, respectively) than 20-nm Au (+27%) and TiN NPs (+35%) as shown from the chopped photocurrent data (Figure 2c) taken at 1.23 V vs. RHE, i.e. the thermodynamic water splitting potential. The observed trend is due to the higher absorption efficiency[29] and corresponding hot electron generation (Supplementary Figure 9) for the case of 50-nm particles compared to 20-nm NPs for both Au and TiN.

To demonstrate the plasmonic origin of the increased performance, we perform wavelength-dependent photocurrent measurements. Incident photon-to-current efficiency (IPCE) plots (Supplementary Figure 10) highlight an enhanced efficiency for TiN and Au plasmonic electrodes in the visible and NIR range, where TiO$_2$ does not absorb photons. When enlarging the plot in the region from 450 to 850 nm, it is clear that the shape of IPCE % curves resembles the line shape of absorption spectra[30,31] of both 20 nm (Figure 3a) and 50 nm (Figure 3b) plasmonic Au and TiN NPs (Figures 3c-d). In the 500-700 nm wavelength range, TiO$_2$/Au structures show the typical plasmon-enhanced efficiency due to hot electron injection, peaking at 550 nm for 20 nm Au and at 600 nm for 50 nm Au. In turn, TiO$_2$/TiN structure exhibits a broader plasmonic absorption peak ranging from 550 nm to 850 nm and further. The increased IPCE % in TiO$_2$/TiN peaks around 700 nm and it is more defined and intense for 50 nm TiN NPs compared to 20 nm ones, as also confirmed by the more intense plasmonic absorption (Supplementary Figure 11).

Our simulations show (Supplementary Figure 9) that TiN has higher absorption efficiency than Au NPs, thus enabling more efficient hot carrier production. In addition, TiN nanostructures show broader plasmonic resonances than Au[32], thus covering a wider wavelength range of the solar spectrum at one selected size. TiN is thus an outstanding platform for applications that exploit physical phenomena coming from plasmon decay such as local heating and hot carrier generation[29,33].

Importantly, $TiO_2$/TiN system exhibits an additional band in IPCE % plot at wavelengths below 500 nm, reflecting the ability of $TiO_2$/TiN to collect also cold carriers generated from interband transitions (see imaginary part of TiN permittivity in Figure 3d)[23], a feature not achievable with conventional plasmonic composites.

**Electrical properties of plasmonic junction interfaces**

The collection of electrons generated from TiN interband excitations is possible only if an Ohmic junction with $TiO_2$ is formed, in contrast to the Schottky barrier created in the case of Au/$TiO_2$. To elucidate the nature of the Au/$TiO_2$ and TiN/$TiO_2$ interfaces, we design thin film devices in a simple geometry and test their electrical properties. Titanium and $TiO_2$ pads are fabricated atop thin films of Au and TiN through standard electron beam lithography and electron beam evaporation (Supplementary Figure 12). The current-voltage (I-V) curves are shown in Figures 4a-b. The Au/$TiO_2$ devices show current rectification as expected for a typical Schottky contact, i.e. we call it a Plasmonic Schottky Interface (PSI). We extracted the Schottky barrier height $\phi_B$ by fitting the I-V curves and find an average value of 0.89 V (see Supplementary Information), well in agreement with previous report of Au/$TiO_2$ barrier height (~1 eV)[23]. The formation of a PSI implies that only hot electrons with $E > \phi_B$ produce the plasmonic enhancement of photocurrent during water splitting. Increasing the carrier concentration in the semiconductor reduces the space charge layer width, and to some extent, increases the tunneling probability of electrons across the potential barrier. In this particular case, we extracted donor density (electron) concentration ($N_D$) in the $TiO_2$ NW array by Mott-Schottky measurements, i.e. in electrochemical conditions, (Supplementary Figure 13) and found $N_D = 2.77 \times 10^{19}$ $cm^{-3}$. At this carrier concentration, hot electron injection across the PSI is regulated mostly by thermionic emission over the barrier even if some carriers can tunnel through the upper level of the potential barrier at $E < \phi_B$. In contrast, those carriers excited from 2.3 eV below the Fermi level (from *d*-band to about the Au Fermi level, i.e. cold electrons) will be reflected from the Schottky barrier and do not contribute to photocurrent (Figure 4c)[23].

On the other hand, TiN/TiO$_2$ devices exhibit Ohmic behavior presenting linear I-V characteristics even when low voltage is applied. That proves that transition metal nitrides have higher Fermi level than conventional noble metals, and form Plasmonic Ohmic Interface (POI) with common semiconductors used in photocatalysis and PVs such as TiO$_2$, α-Fe$_2$O$_3$, and Si[34–38]. The formation of a POI in TiN/TiO$_2$ uniquely enables the collection of both hot carriers due to plasmonic excitation and cold electrons due to TiN interband transition (Figure 4d).

Another advantage of forming a POI is to break the limitation $E > \phi_B$ in hot electron collection, thus opening the path to the utilization of hot carriers beyond the NIR (i.e., wavelength range enabled by Au/Ti/Si structures with typical $\phi_B = 0.5$ eV)[13,14] and those hot carriers that lose energy due to electron-electron scattering during their way toward the metal-semiconductor interface thus providing enhanced water splitting performance when compared to noble metal-modified TiO$_2$ electrodes.

The electronic band alignment (Figures 4c-d) for both Schottky and Ohmic devices show that wide bandgap semiconductors, i.e. TiO$_2$, allow preferential collection of electrons. Hot electrons and holes experience truly different barrier heights. For electrons, the Fermi level of the plasmonic material is forced to align with the conduction band of TiO$_2$, resulting in a very large potential barrier for holes[23,17]. One can also argue that Ohmic junctions in plasmonic hot carrier devices would allow charge flow in both directions thus mitigating the benefits of using resonant materials. However, in solar energy conversion applications such as PVs, photoelectrochemical water splitting, and CO$_2$ reduction, at least a single semiconductor capable of photovoltage generation (or the application of an external bias) favors the hot electron collection on one side of the device (Supplementary Figure 14). The absence of a potential barrier induces large dark currents in photodetection applications. However, by engineering the POI with an additional layer[13,14] it is possible to create a tailored barrier so that noise due to thermal excitation is minimized.

We consider a simple physical model[15] that describes hot electron collection across a metal-semiconductor (e.g. plasmonic) interface to show that TiN offers enhanced performance compared to Au

NPs. The model comprises three fundamental steps: hot electron generation (1), hot electron migration towards the plasmonic interface (2), and hot electron injection into the conduction band of the TiO$_2$ NWs (3)[15]. Firstly, in addition to a higher generation rate of hot electrons (1), TiN assures an increased number of hot electrons reaching the plasmonic interface (2) due to (i) a 30% longer effective mean free path compared to Au[40,41], (ii) a lower carrier concentration than Au[39] that diminishes the energy loss of hot carriers due to inelastic scattering, (iii) the TiN/TiO$_2$ Ohmic contact that facilitates the additional collection of those electrons that lose energy near the plasmonic interface. Finally, during hot electron injection (3), carriers typically acquire low transmission probability because of considerable momentum mismatch and poor overlap of the electron wavefunctions between the metal and the semiconductor[15]. While momentum conservation is broken when the dimension of the plasmonic component is much lower than the photon wavelength[15,42–44], as in the case of NPs, the overlap of electronic wavefunctions still strongly influences electron injection efficiency. In our photocatalytic system hot electrons produced in TiN have a stronger overlap of their wavefunctions with TiO$_2$ when compared to Au, since the density of states of conduction band is mainly characterized by Ti 3d wavefunctions for both TiN and TiO$_2$[39,45–47]. The overlap of wavefunctions enhances hot electron transmission across the interface and holds potential for other plasmonic transition metal nitrides such as TaN[18] for the efficient coupling with its semiconducting polymorph Ta$_3$N$_5$, i.e. one of the most active photoanodes[48] for PEC water splitting ever reported.

In summary, plasmonic TiN NPs on TiO$_2$ nanowires produce 25% higher photocurrent enhancement in solar water splitting when compared to their Au counterpart. We show that the improved performance achieved with TiN is due to the higher absorption efficiency in a broader spectral region that leads to enhanced hot electron generation. In contrast to Au, TiN forms an Ohmic junction with TiO$_2$ that uniquely enables the collection of cold electrons from TiN interband transitions, enhances electron mean free path and hot electron transmission across the interface. Consequently, TiN-TiO$_2$ structure provides more efficient hot electrons collection across a plasmonic interface. The "downhill" collection of hot carriers breaks the limitations of traditional Schottky contacts and enables solar harvesting beyond the NIR region. These results can be extended to other plasmonic metal nitrides and alternative plasmonic materials which have the Fermi level inside their conduction band, i.e. highly-doped transparent conducting oxides. Our approach marks a departure from conventional plasmonic metals into new material platforms for solar energy-powered devices with improved performance by engineering novel hybrid metallic/semiconducting building blocks with different electronic and interfacial properties that can be created and assembled on demand.

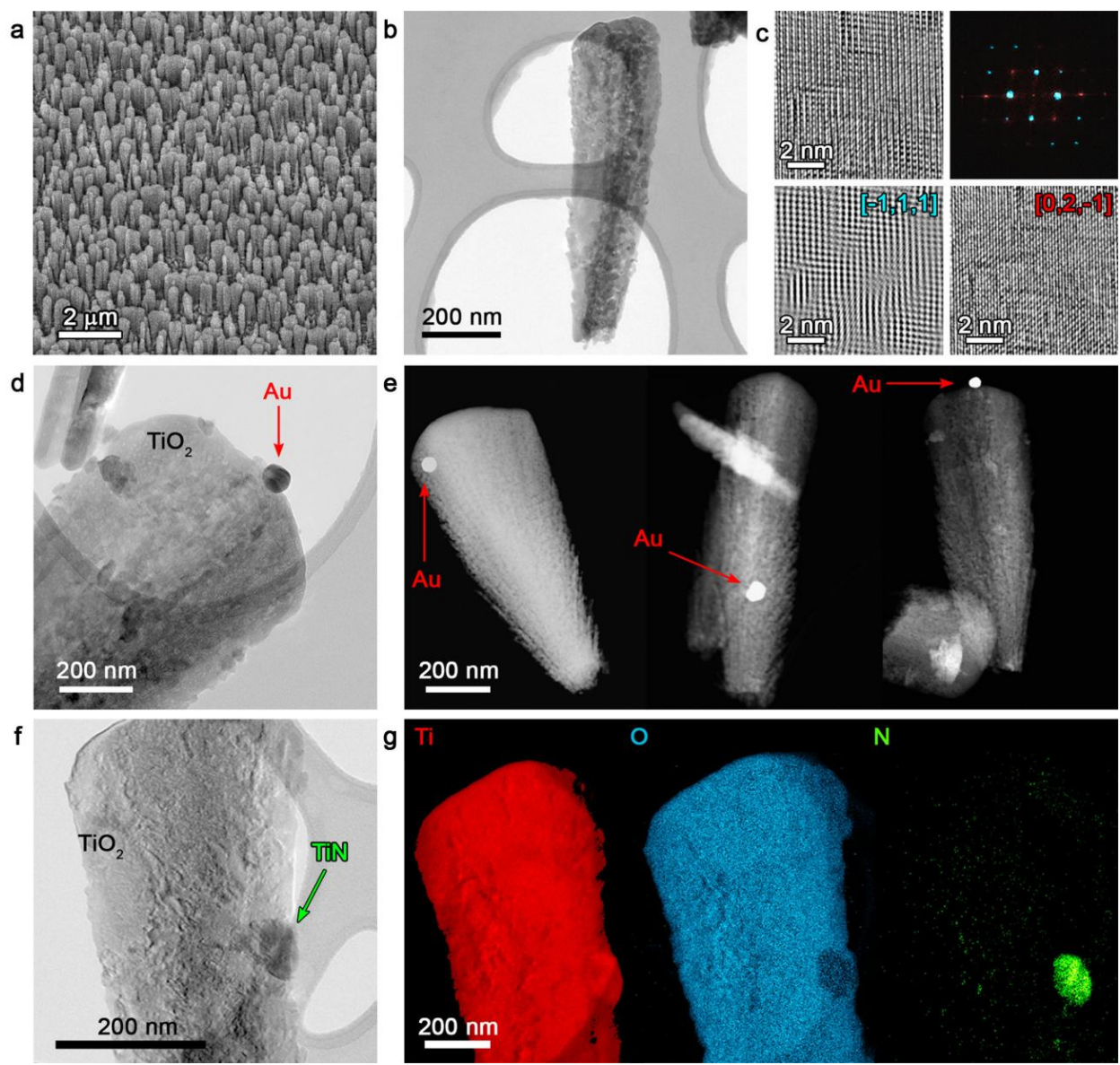

**Figure 1. Morphology of bare and NP-loaded TiO$_2$ nanowire array. a,** SEM image showing the large-scale homogeneity of TiO$_2$ nanowires grown through GLAD. **b,** TEM micrograph of a single TiO$_2$ nanowire. **c,** HRTEM of selected portion of **b** and related FFT analysis and crystalline indexing along [-1,1,1] and [0,2,-1] zone axis. **d,** TEM image of an Au NP sitting on the top of a single TiO$_2$ nanowire. **e,** HAADF-STEM images showing Au NPs at different locations of TiO$_2$ nanowire. **f,** TEM image of a TiN nanocube on the lateral surface of a TiO$_2$ nanowire. **g,** Electron spectroscopy images (ESI) map for Ti (red), O (blue), and N (green) for TiN/TiO$_2$ composite, respectively.

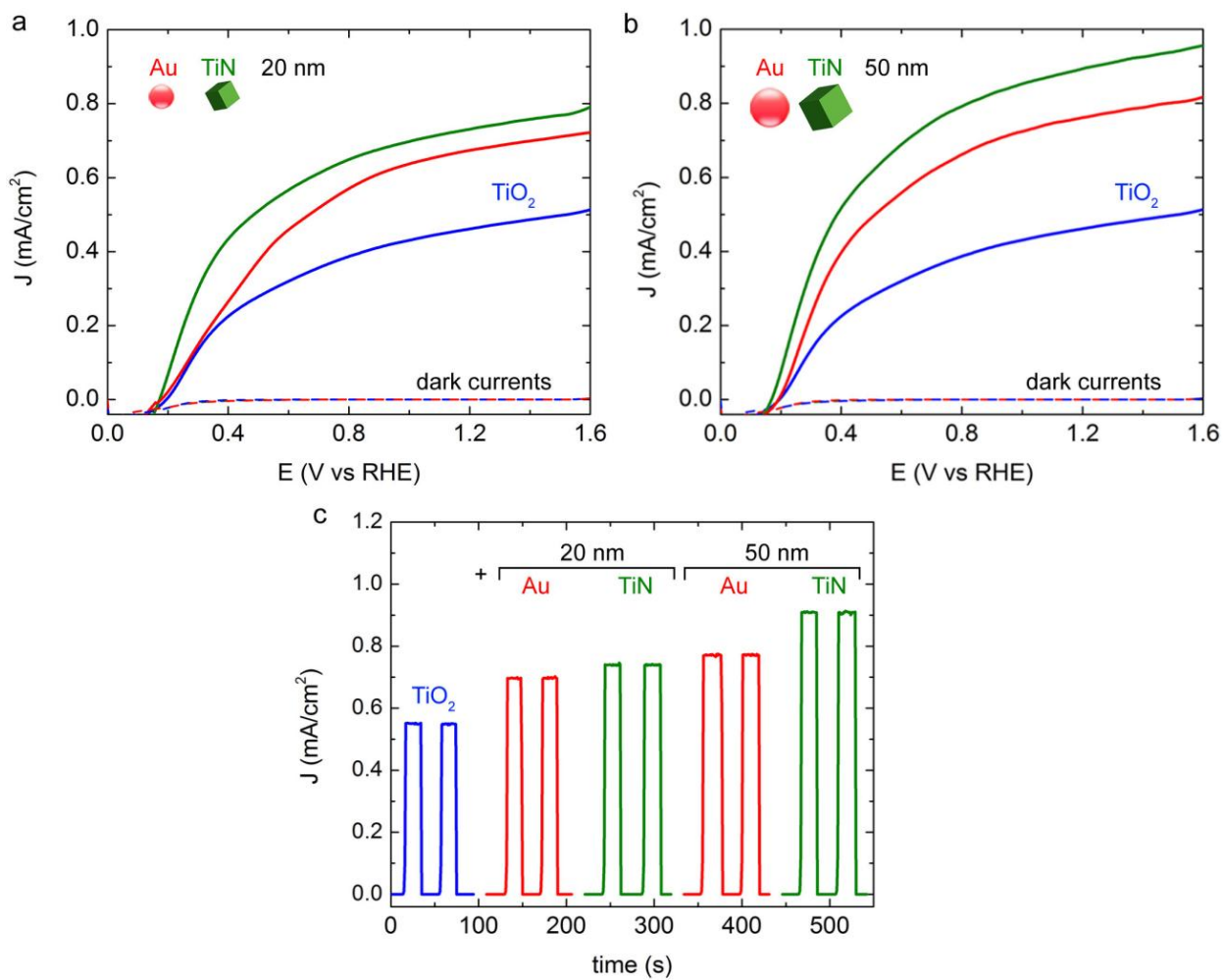

**Figure 2. Photocurrent density-applied potential (J-V) curves and incident photon-to-current efficiency (IPCE). a,** Anodic photocurrent for $TiO_2$ NWs, $TiO_2$ NWs –Au 20 nm, $TiO_2$ NWs –TiN 20 nm. **b,** J-V plot for $TiO_2$ NWs, $TiO_2$ NWs –Au 50 nm, $TiO_2$ NWs –TiN 50 nm. **c,** Chopped photocurrent for all samples at 1.23 V vs reversible hydrogen electrode (RHE). Color legend is $TiO_2$ (blue line), $TiO_2$ – Au (red line), $TiO_2$ –TiN (green line). All measurements are taken under 150 mW solar irradiation.

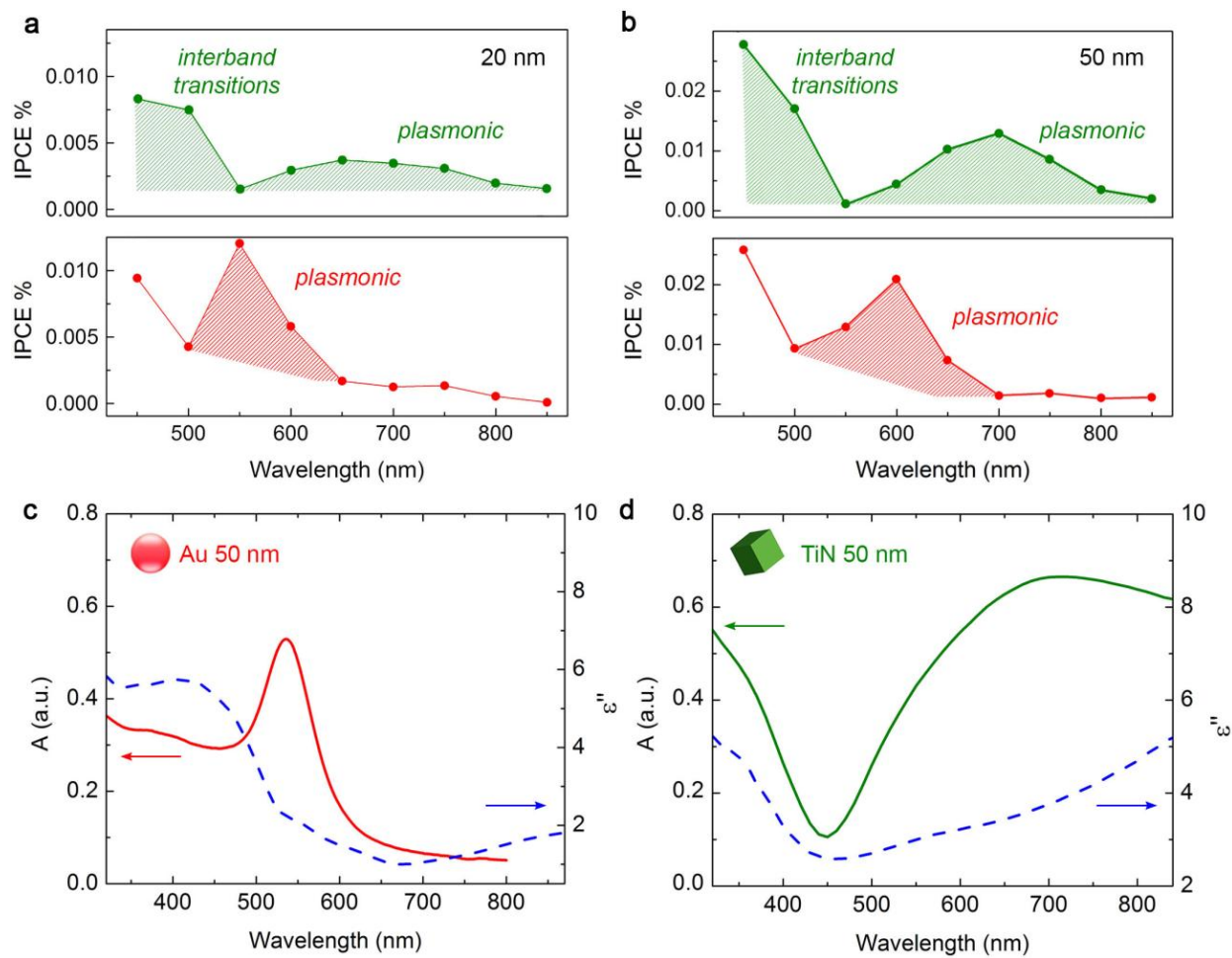

**Figure 3. Incident photon-to-current efficiency (IPCE) and optical properties.** Wavelength-dependent IPCE % showing the enhancement due to plasmonic particles for **a,** TiO$_2$ NWs – Au 20 nm (red line), TiO$_2$ NWs – TiN 20 nm (green line), and **b,** TiO$_2$ NWs –Au 50 nm (red line), TiO$_2$ NWs –TiN 50 nm (green line). Absorption spectrum and imaginary part of permittivity ($\varepsilon''$) for **c,** Au NPs (50 nm) and **d,** TiN nanocubes (50 nm) in solution[31]. Permittivity values of gold are taken from Johnson and Christy[30].

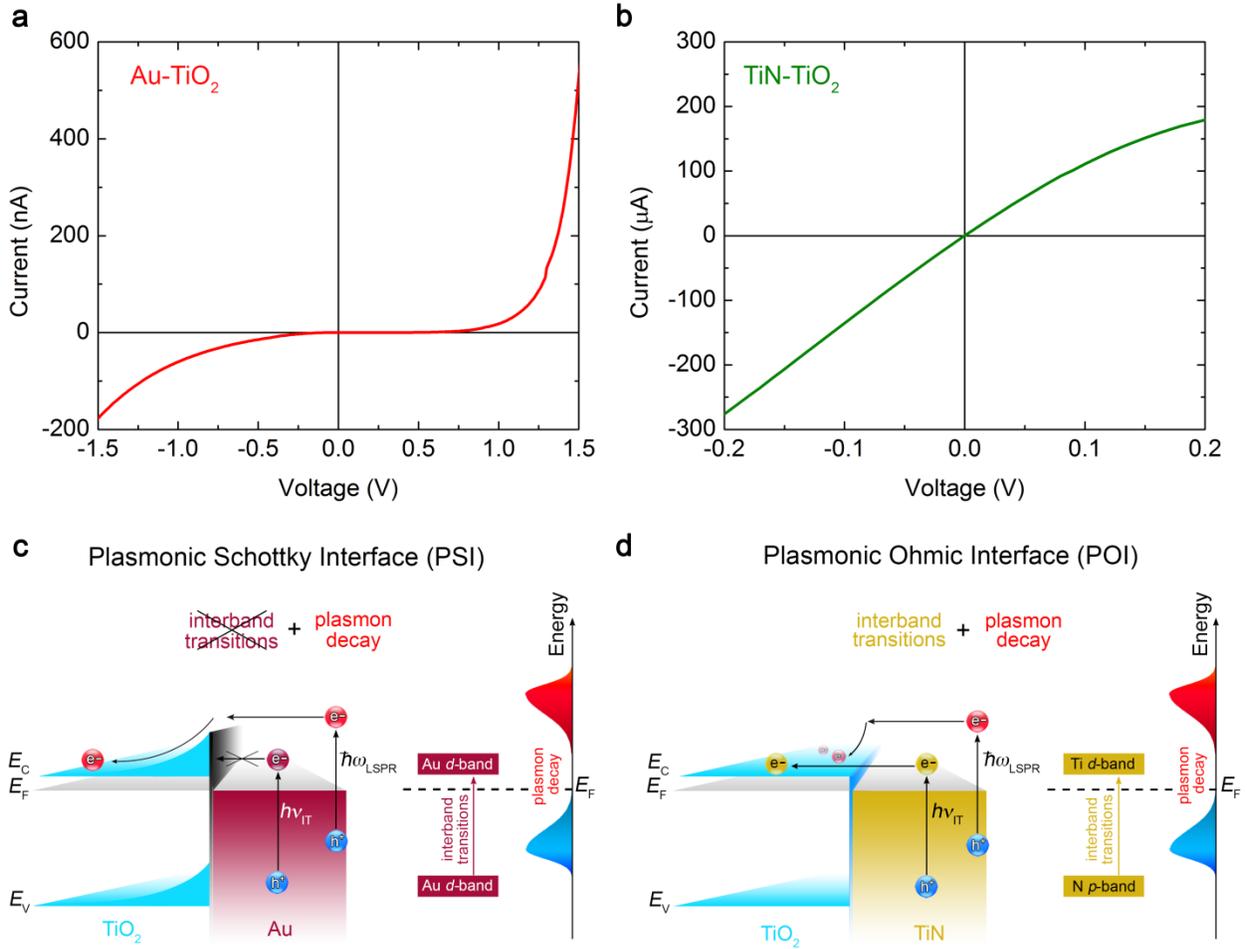

**Figure 4. Electrical measurements and band diagram of the plasmonic interfaces. a,** Current-voltage (I-V) curve of a Au-TiO$_2$ Schottky device. **b,** I-V curve of a TiN-TiO$_2$ Ohmic device. **c,** Band diagram of a Plasmonic Schottky Interface (PSI). **d,** Band diagram of a Plasmonic Ohmic Interface (POI). The PSI prevent the collection of "cold" carriers excited from interband transitions and hot carriers with E < $\phi_B$ (0.89 eV). The POI has no potential barrier and allows high photocurrent due to both hot electrons and "cold" electrons flow.

**Methods**

**Growth of TiO₂ nanowires and decoration with plasmonic nanoparticles.** The $TiO_2$ nanowire films on an FTO substrate are fabricated using the glancing angle deposition (GLAD) technique. The substrate normal is tilted at 86° with respect to the deposition source. The resulting films are crystallized in the anatase phase by annealing at 500 °C for 6 h. To decorate the $TiO_2$ nanowires with plasmonic NPs, $TiO_2$ nanowire thin films grown on FTO substrate are vertically immersed into an Au or TiN NP solution for 24 h. The samples are then thoroughly washed by ultrapurified water and dried under a nitrogen flow. Citrate-stabilized Au NPs with 20 and 50-nm diameters were purchased from Sigma Aldrich. TiN powders with average crystal dimensions of 20 and 50 nm were purchased from PlasmaChem.

**Characterization.** X-ray diffraction (XRD) patterns are recorded at room temperature in the $30° \leq 2\theta \leq 105°$ range employing the Cu-K$\alpha$ radiation. Rietveld analysis is performed subtracting the background using the shifted Chebyshev polynomials, and the diffraction peak profiles are fitted with a modified pseudo-Voigt function. TiN crystal dimensions are determined from the Sherrer equation. SEM images of the $TiO_2$ nanowires arrays are taken with a Hitachi S-4800 Field Emission SEM. Transmission electron microscopy (TEM and HRTEM) analysis are performed by a ZEISS LIBRA200FE. Electron Spectroscopic Imaging (ESI) and elemental maps are collected via energy-filtering electrons at related Ti $L_2$ edge (449 eV) O K edge (532 eV) and N K edge (401 eV) of the electron energy loss spectrum (EELS). The ZEISS in-column Omega filter spectrometer is used for all the EELS measurements, coupled whit the Olympus-SIS iTEM software for data collection and analysis. TEM specimens are collected by scratching the surface and collecting the fragments which adherence onto a holey carbon supported film on Cu 300 mesh grid. Transmission spectra of Au and TiN NPs are taken by using a PerkinElmer Lambda950 UV-VIS-NIR spectrophotometer.

**Photoelectrochemical measurements.** Photoelectrochemical characterization is carried out in a standard three electrode cell with a PGSTAT204 Autolab potentiostat. A high surface area Pt mesh is the counter electrode. The applied potential ($E$) is referred to the Ag/AgCl electrode scaled to the reversible hydrogen electrode (RHE) through the Nernst equation:

$E_{RHE} = E_{AgCl} + 0.197 \, \text{V} + 0.059 \, p\text{H}$

where $E_{AgCl}$ is the measured applied potential and 0.197 is the Ag/AgCl standard potential vs. the normal hydrogen electrode. We report all measurements with respect to RHE. The measurements are carried out in 1 M NaOH aqueous solution at pH 13.6. *J-V* curves are measured at a scan rate of 10 mV/s.

The incident photon-to-current efficiency (IPCE) measurements are carried out by applying a constant bias (1.23 $V_{RHE}$). IPCE are calculated using the following equation:

$$IPCE\ (\%) = \frac{J_{photo} \times 1240}{J_0 \times A \times \lambda} \times 100$$

where $J_{photo}$ (mA/cm$^2$) is the differential (light-dark) measured photocurrent for TiO$_2$ nanowires film, $J_0$ (mA/cm$^2$) is the measured power density of the light source at a specific wavelength, and A (cm$^2$) is the geometrical area of the electrode. Wavelength selection is possible by applying a set of band pass filter (FWHM = 40 nm, Thorlabs) at the light source. A 300 W Xenon arc lamp (calibrated at 150 mW/cm$^2$) is used as the light source. At least five electrodes of each type are fabricated and tested. All electrodes show similar characteristics and representative data are reported.

**I-V curve measurements.** To measure the electrical properties of the TiO$_2$-TiN (TiO$_2$-Au) junction, a 50 × 50 μm patch of 100-nm TiO$_2$ with an 80-nm Ti cover is fabricated on top of the TiN (Au) 100-nm thick films by standard electron beam lithography and electron beam deposition. The TiN film is deposited with DC magnetron sputtering. The Au substrate is deposited with e-beam evaporation. The current-voltage (I-V) curve are obtained by applying voltage between TiN (Au) film (drain) and Ti (source) patch. Measurements are repeated at least three times.

**Acknowledgements**

We acknowledge financial support from the U.S. National Science Foundation (MRSEC program), grant number DMR-1120923 and DMR-506775; the Italian Ministry of Education, University and Research (MIUR) through the FIRB project "Low-cost photoelectrodes architectures based on the redox cascade principle for artificial photosynthesis" (RBFR13XLJ9). The authors thank Yuqi Zhu for assistance in the I-V curve measurement and N. Kinsey for discussion on the manuscript.

**Supplementary Information**

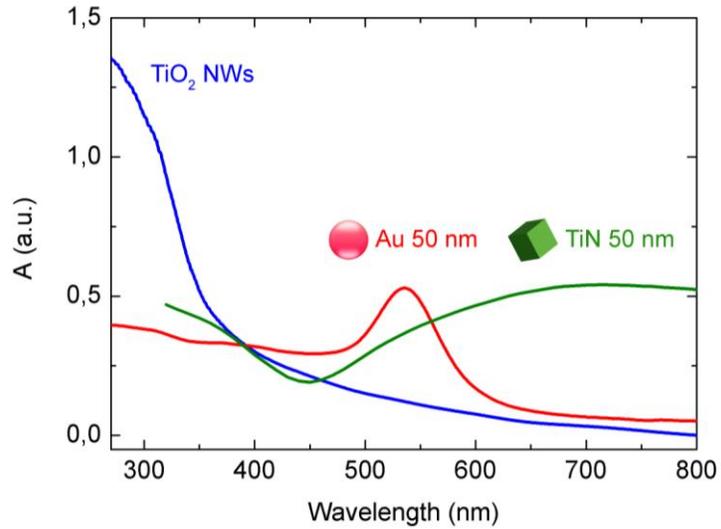

**Supplementary Figure 1.** Experimental absorption spectra of $TiO_2$ NWs (blue line), Au nanoparticles (red line) and TiN nanocubes (green line) 50 nm in average size.

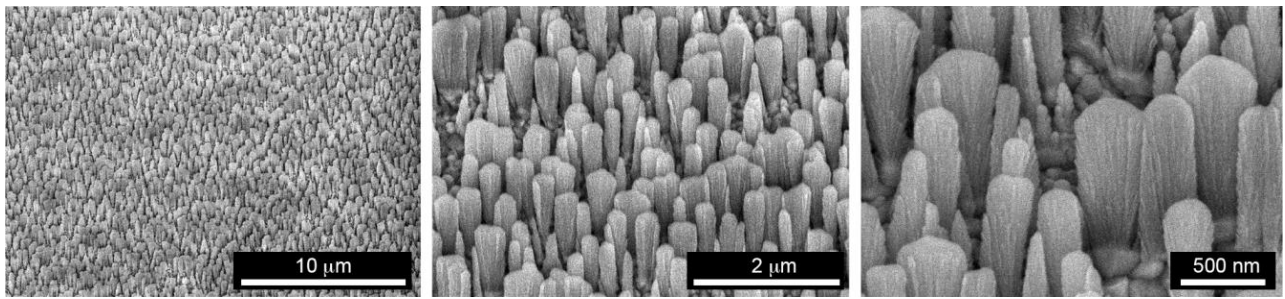

**Supplementary Figure 2.** SEM images of $TiO_2$ nanowires grown on FTO glasses by using Glancing Angle Deposition (GLAD).

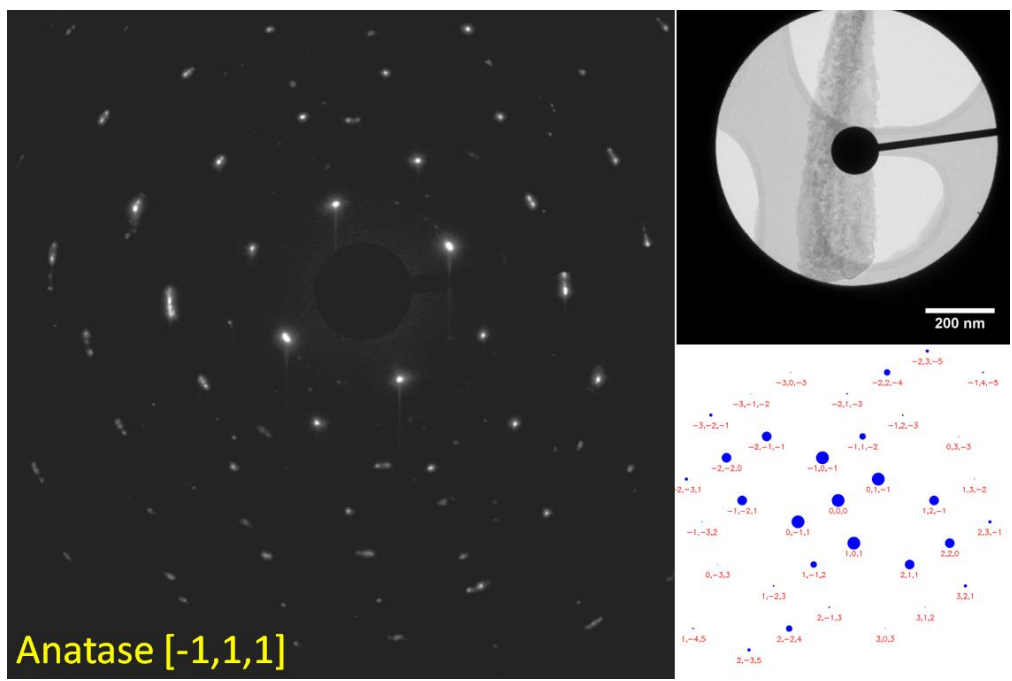

**Supplementary Figure 3.** Selected Area Electron Diffraction (SAED, left side) of a single $TiO_2$ nanowiwe (upper right side) showed a recognized and characteristic pattern for anatase along zone axis [-1,1,1]. Diffraction indexing was performed with the Diffraction Image Analysis module of iTEM Software (Olympus SIS). The result is in line with FFT local analysis performed on selected high resolution micrographs and with Web-EMAPS simulation (lower right side)[1].

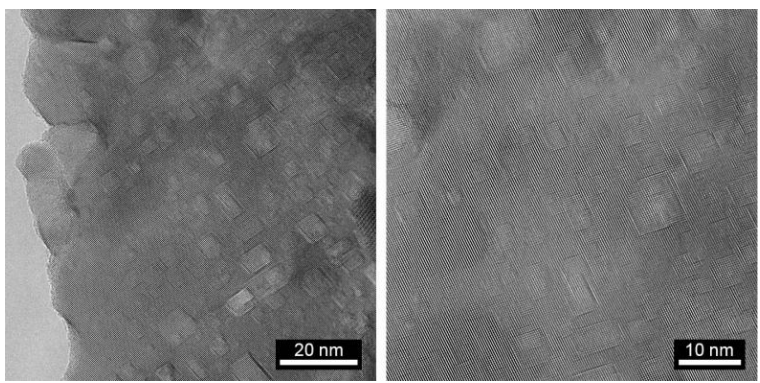

**Supplementary Figure 4.** HRTEM images of a portion of single $TiO_2$ NW highlighting the high crystal orientation of GLAD structures. The majority of the $TiO_2$ lattice fringes have the same orientation, while some area showing orthogonal and complex lattice fringes indicate the growth of a secondary structure into the single crystal structure of TiO$_2$ NWs.

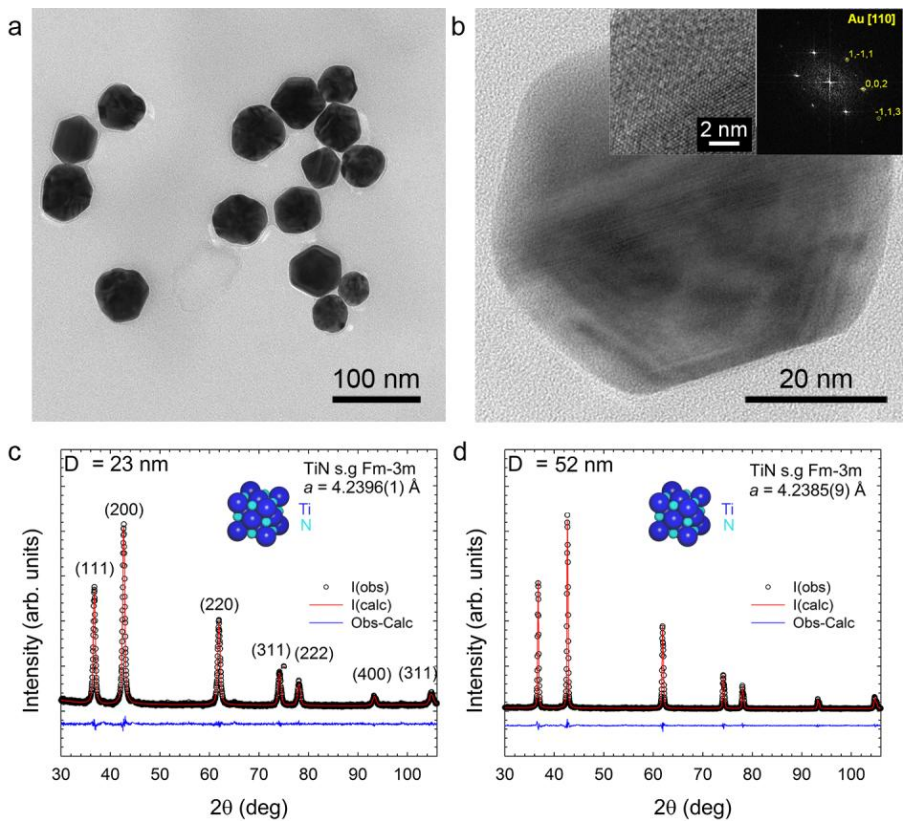

**Supplementary Figure 5. a-b,** TEM micrographs of Au nanoparticles with size distribution centered at 50 nm. Inset of b shows HRTEM and related FFT analysis of a single Au nanoparticle. **c-d,** X-ray powder diffraction patterns of TiN 20 nm (c) and 50 nm (d). Average crystal sizes obtained from Rietveld refinement are 23 and 52 nm, while the lattice constant of both sample approached the tabulated bulk value of 4.24 Å.

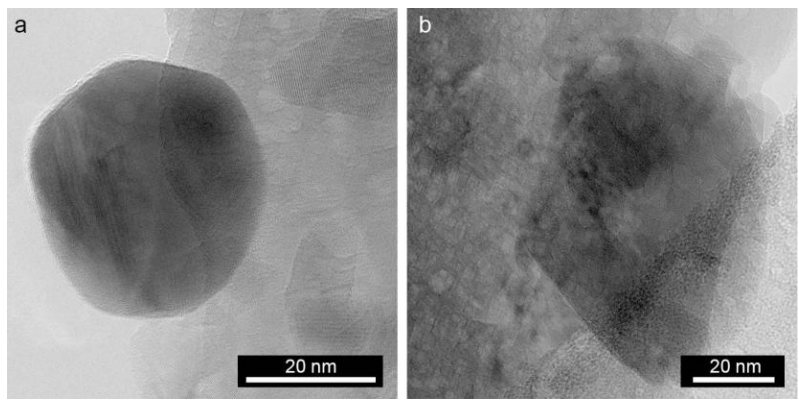

**Supplementary Figure 6.** HRTEM images showing a single Au (**a**) and TiN (**b**) nanoparticle attached on a TiO$_2$ NW surface. Large darker area in the micrographs represent Au and TiN nanoparticles. The TiN nanocube shows a higher contact area than the Au nanoparticle.

**Electron Energy Loss Spectroscopy (EELS) analysis**

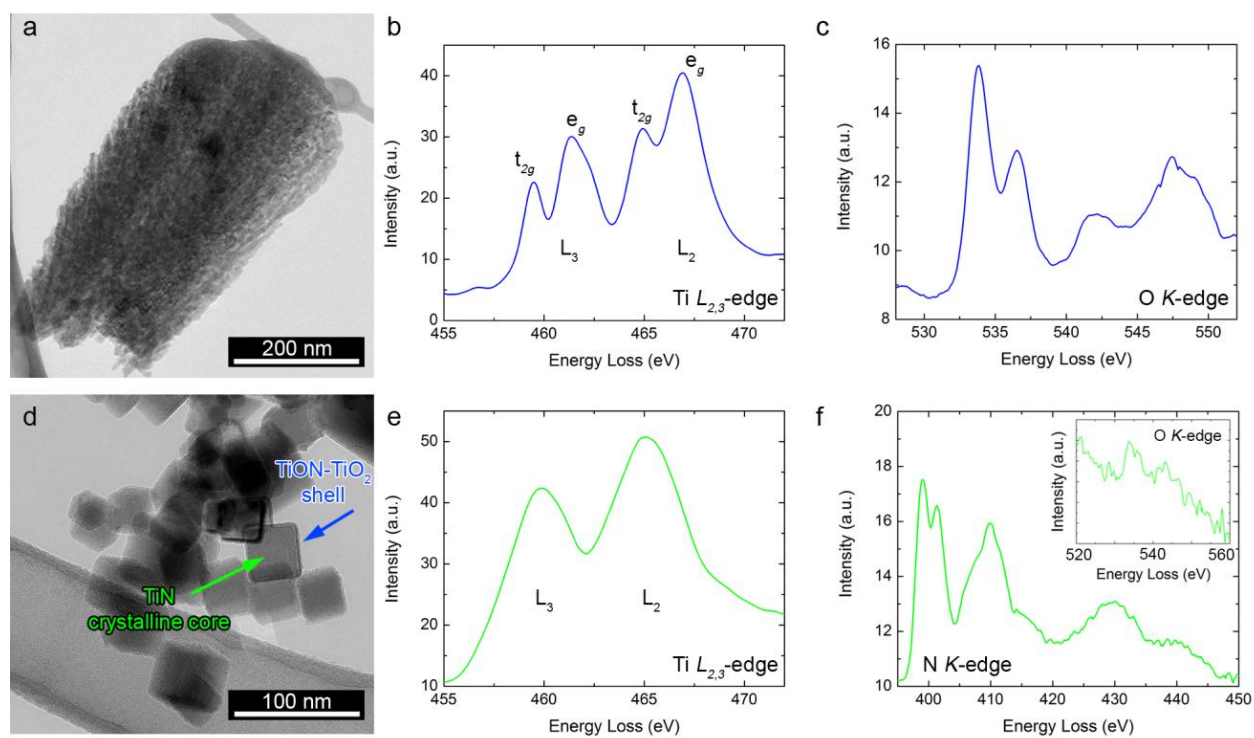

**Supplementary Figure 7.** TEM image (**a**) and related Electron Energy Loss Near Edge Structure (ELNES) spectra at the Ti $L_{2,3}$-edge (**b**) and O $K$-edge (**c**) for of a single TiO$_2$ NWs. TEM image (**d**) and related ELNES spectra at the Ti $L_{2,3}$-edge (**e**), N $K$-edge (**f**) and O $K$-edge (**f**, inset) of TiN nanocubes (see ref. 6 for a more detailed discussion).

**Supplementary Table 1.** ELNES spectra analysis at the Ti $L_{2,3}$-edge for TiO$_2$ NWs and TiN nanocubes.

| Sample | Ti $L_3$ (eV) | | Ti $L_2$ (eV) | | $L_2 - L_3$ (eV) | $L_3 (e_g) - L_3 (t_{2g})$ (eV) | $L_2 (e_g) - L_3 (t_{2g})$ (eV) |
|---|---|---|---|---|---|---|---|
| **TiO$_2$** | $t_{2g}$ | 459.5 | $t_{2g}$ | 464.9 | 5.4 | 1.9 | 2 |
|  | $e_g$ | 461.4 | $e_g$ | 466.9 | 5.5 |  |  |
| **TiN** |  | 459.8 |  | 465.1 | 5.2 | 1.3 | 1.3 |

ELNES appears above the absorption edge in the electron energy loss spectrum (EELS) and allows a qualitative, and sometime quantitative, interpretation of unoccupied electronic states. EELS maps empty states above the Fermi level thus allowing us to probe the density of states in the conduction band of the materials. Ti–$L$ and O–$K$ edges are the main features present in the TiO$_2$ NWs spectrum (Supplementary Figure 6b-c, respectively). The Ti $L_{2;3}$ edge shown in Supplementary Figure 7b has the typical shape of the anatase TiO$_2$ phase and mainly reflects Ti $3d$ unoccupied states split into the $t_{2g}$ and $e_g$ sub–bands because the octahedral coordination of Ti atoms with O[2,3]. The Ti $L_{2,3}$ edge of TiN nanocubes (Supplementary Figure 6e) does not show this splitting due to the different crystal coordination (Supplementary Table 1). The first two peaks observed in the ELNES of Supplementary Figure 6f can be attributed to the transitions from the 1s state to the unoccupied nitrogen 2p states hybridized with titanium 3d states with symmetries of $t_{2g}$ and $e_g$, respectively. The peaks are very well separated and the second peak has a lower intensity than the first one, indicating the absence of nitrogen vacancies in the TiN nanocubes and confirming their high single-crystal quality[4,5]. TiN EELS spectrum shows also a low intensity signal in the energy range related to O K-edge (Supplementary Figure 6f, inset) due to both TiO$_2$ and TiON species present in the shell of nanocubes[6].

**Open-circuit voltage measurements**

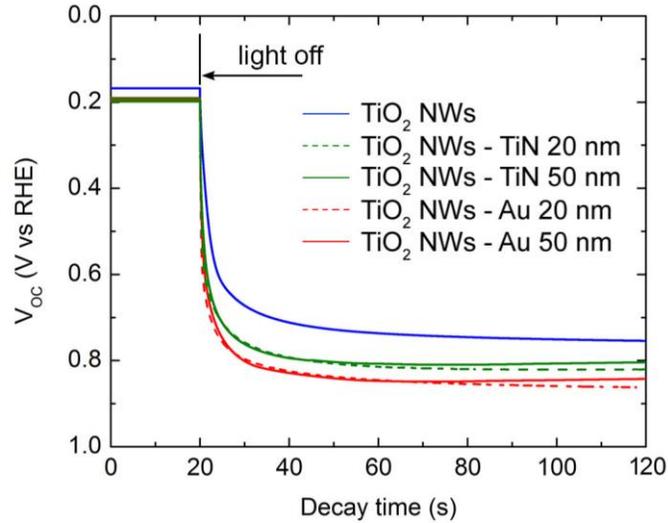

**Supplementary Figure 8.** Open circuit voltage ($V_{OC}$) decay of TiO$_2$ NWs (blue line), TiO$_2$ NWs – Au 20 nm (red solid line), TiO$_2$ NWs – Au 50 nm (red dashed line), TiO$_2$ NWs – TiN 20 nm (green solid line), TiO$_2$ NWs – TiN 50 nm (green dashed line).

Open-circuit voltage ($V_{OC}$) measurements (Supplementary Figure 7) are carried out in a standard three electrode electrochemical cell where the TiO$_2$ NW films grown on FTO were the working electrodes. $V_{OC}$ decay measurements give access to the determination of the quasi-Fermi level of the TiO$_2$ NWs before and after the decoration with plasmonic nanoparticles in light and dark conditions (Supplementary Table 2).

$V_{OC}$ decay measurement is a useful method to examine the passivation of surface traps in a photoelectrode material and consequently to compare the coverage of Au and TiN nanoparticles on TiO$_2$ NWs. We calculated the decay lifetime of each $V_{OC}$ – t profiles by fitting to a biexponential function with two time constants:

$$V_{OC}(t) = A_0 + A_1 e^{-\frac{t}{\tau_1}} + A_2 e^{-\frac{t}{\tau_2}}$$

$$\tau_m = (\tau_1 \tau_2) / (\tau_1 + \tau_2)$$

Where $\tau_m$ is the harmonic mean of the lifetime and the total half life time ($\tau$) is $\log(2 \times \tau_m)$[7].

**Supplementary Table 2.** Open circuit voltage in light ($V_{OC}$ light) and dark ($V_{OC}$ dark) conditions along with device characteristic photovoltage ($V_{ph} = V_{OC}^{light} - V_{OC}^{dark}$) values, and half life time ($\tau$) for all photoanodes.

| Sample | $V_{OC}$ (light) ($V_{RHE}$) | $V_{OC}$ (dark) ($V_{RHE}$) | $V_{ph}$ (V) | $\tau$ (sec) |
|---|---|---|---|---|
| **TiO$_2$ NWs** | +0.170 ± 0.010 | +0.750 ± 0.030 | 0.580 | 0.54 |
| **TiO$_2$ NWs - Au 20 nm** | +0.200 ± 0.020 | +0.840 ± 0.020 | 0.640 | 0.15 |
| **TiO$_2$ NWs - Au 50 nm** | +0.190 ± 0.010 | +0.860 ± 0.030 | 0.670 | 0.22 |
| **TiO$_2$ NWs - TiN 20 nm** | +0.200 ± 0.020 | +0.810 ± 0.030 | 0.610 | 0.21 |
| **TiO$_2$ NWs - TiN 50 nm** | +0.190 ± 0.020 | +0.820 ± 0.020 | 0.630 | 0.14 |

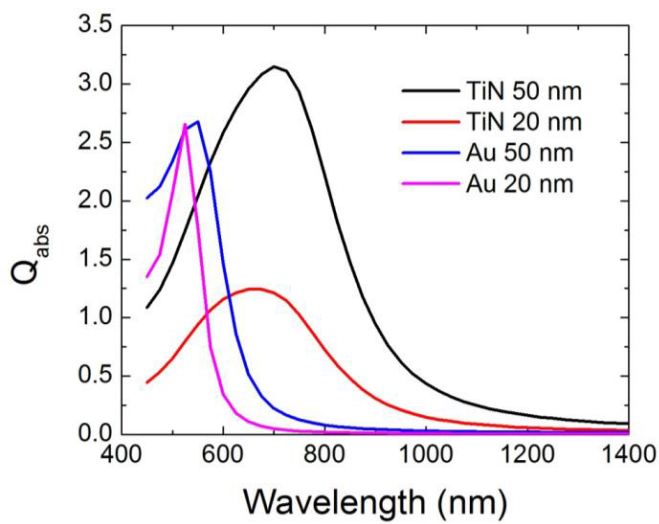

**Supplementary Figure 9.** Simulated absorption efficiency of Au nanoparticles and TiN nanocubes with 20 and 50 nm size in water. Calculations were performed with a commercial software (COMSOL Multiphysics) based on finite element method.

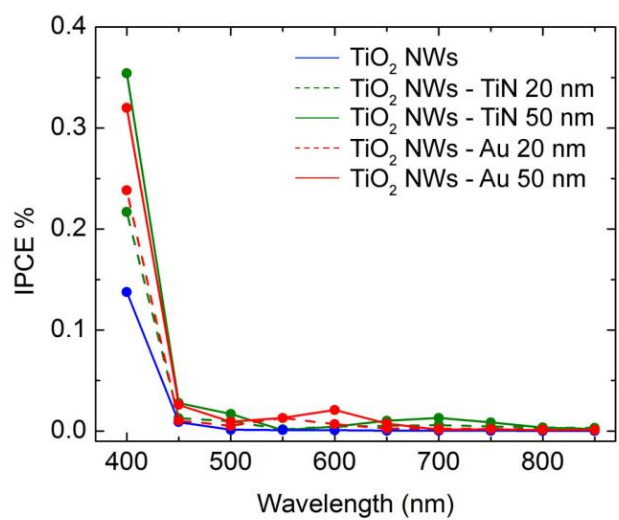

**Supplementary Figure 10.** IPCE % for all photoanodes recorded at 1.23 V vs RHE. Color legend is $TiO_2$ (blue line), $TiO_2$ – Au (red line), $TiO_2$ –TiN (green line). All measurements are taken under 150 mW solar irradiation.

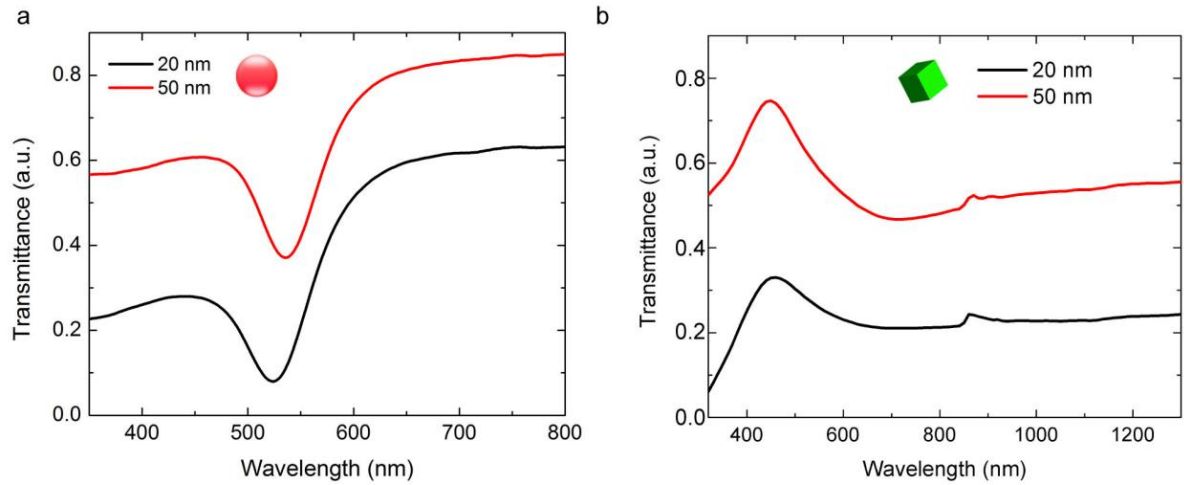

**Supplementary Figure 11.** Experimental transmittance spectra of (a) Au nanoparticles and (b) TiN nanocubes dispersed in aqueous solution.

**Electrical measurements and discussion on the metal-semiconductor interface**

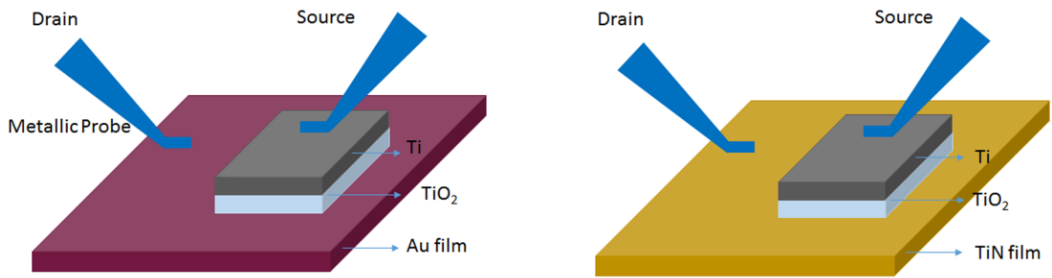

**Supplementary Figure 12.** Schematic showing the device geometry adopted for I-V curve measurements on Au-TiO$_2$ and TiN-TiO$_2$ samples.

The I-V curve for the Au-TiO$_2$ Schottky device is fitted by using the diode equation as follows:

$$V = \frac{nkT}{e} \ln\left(\frac{J + J_0}{J_0}\right) + JRA$$

$$J = A^* T^2 \exp\frac{q\,\emptyset_B}{kT}$$

Fitted parameters[8]: Schottky barrier, $\emptyset_B = 0.885$ eV; Richardson constant: A*=1200; ideality factor: n=5; series resistance: R=2000 Ω;

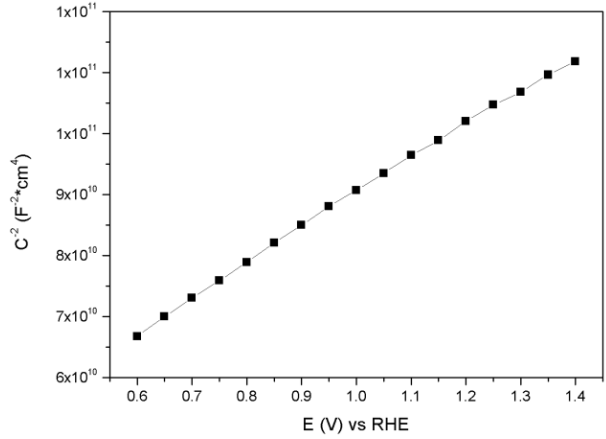

**Supplementary Figure 13.** Mott−Schottky plot for TiO$_2$ NW array photoelectrode obtained in the dark in a standard three electrochemical cell.

From the Mott−Schottky equation, the donor concentration ($N_D$) can be extracted by:

$$\left(\frac{A_s}{C_{bulk}}\right)^2 = \frac{2}{q\varepsilon_r\varepsilon_0 N_D}\left(V - E_{FB} - \frac{k_B T}{q}\right)$$

Where $C_{Bulk}$ is the space charge capacitance, $A_s$ is the geometrical area of the electrode (1 cm$^2$), $V$ is the applied potential, $E_{FB}$ is the flat band potential of the semiconductor, $k_B$ is the constant of Boltzmann (1.38 10$^{-23}$ J K$^{-1}$), $T$ is the temperature (298 K), $q$ is the electron charge (1.602 10$^{-19}$ C), $\varepsilon_0$ is the permittivity under vacuum (8.85 10$^{-12}$ C$^2$ J$^{-1}$ m$^{-1}$) and $\varepsilon_r$ is the dielectric constant (55 for TiO$_2$ anatase)[9]. The calculated value for the TiO$_2$ NW array is $N_D = 2.77 \times 10^{19}$ cm$^{-3}$.

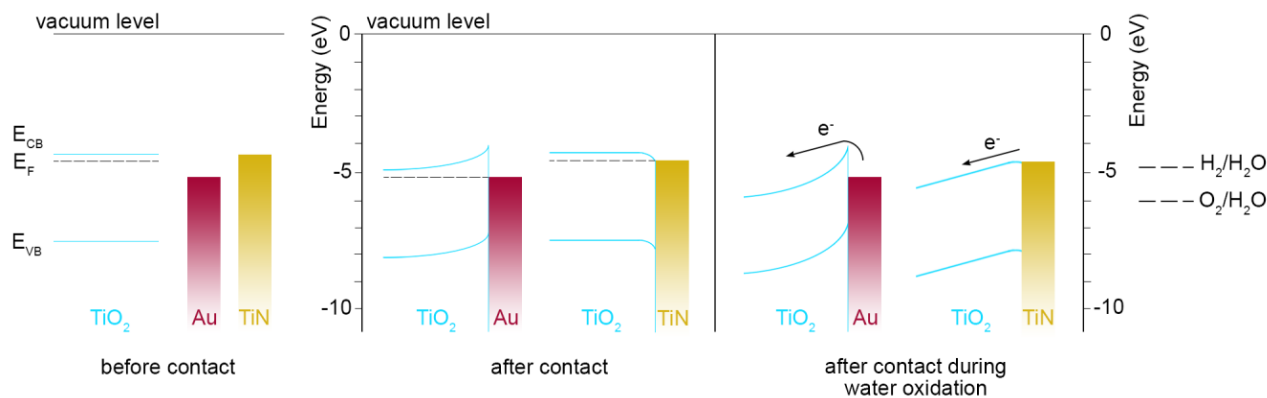

**Supplementary Figure 14.** Band diagram showing the electronic alignments between $TiO_2$, TiN, and Au before contact, after interfacial contact, and during water oxidation, i.e., under an anodic bias. In these conditions, hot electron collection is downhill and current is observed only in one direction also for Ohmic junction devices.

**Supplementary References**

1. Zuo, J. M. & Mabon, J. C. Web-based Electron Microscopy Application Software: Web-EMAPS. *Microsc. Microanal.* **10** (2), 2004.

2. Naldoni, A. *et al.* Influence of $TiO_2$ electronic structure and strong metal–support interaction on plasmonic Au photocatalytic oxidations. *Catal. Sci. Technol.* **6**, 3220-3229 (2016).

3. Naldoni, A. *et al.* Effect of nature and location of defects on bandgap narrowing in black $TiO_2$ nanoparticles. *J. Am. Chem. Soc.* **134**, 7600–7603 (2012).

4. Woltersdorf, J., Feldhoff, A., Lichtenberger O. The complex bonding of titanium nitride layers in C/Mg composites revealed by ELNES features. *Cryst. Res. Technol.* **35**, 653–661 (2000).

5. Tsujimoto, M. *et al.* Influence of nitrogen vacancies on the N K-ELNES spectrum of titanium nitride. *J. Electron Spectrosc.* **143**, 159–165 (2005).

6. Guler, U. *et al.* Colloidal plasmonic titanium nitride nanoparticles: properties and applications. *Nanophotonics* **4**, 269–276 (2015).